# [1]Wave characteristics and anisotropic homogenization theory of soft matters layered structure

Rui.Guo[a], Kai Zhang[a,b,*], Nicholas X. Fang [c*]

[a] *School of Aerospace Engineering, Beijing Institute of Technology, Beijing, 100081, China*

[b] *Tangshan Research Institute Beijing Institute of Technology, Tangshan, 063000, China*

[c] *Department of Mechanical Engineering, The University of Hong Kong, Hong Kong, Special Administrative Region of China*

[*] Corresponding author. E-mail: zhangkai@bit.edu.cn; nicxfang@hku.hk

## Abstract

We investigate in this work the wave characteristics and homogenization theory of soft matter layered structure in the limit of low-frequency P-wave. Using the method of potentials, we derive closed-form dispersion relationship and identify three distinct modes of the soft matter layered structure: quasistatic mode, resonance mode, and slip mode. These modes differ based on their equivalent interface conditions: a continuous interface for quasistatic mode, a spring-like interface for resonance mode, and a slip-like interface for slip mode. Additionally, we propose a simplified model capturing P-wave wave characteristics in the S-wave high-frequency regime. Our findings unify wave-structure relationships across solid, liquid, and soft matter composites, offering a predictive framework for engineering metamaterials with programmable wave responses. This study offers new insight on the fundamental understanding of layered media and providing direct design principles for applications in acoustic cloaking, vibration damping, and biomedical imaging.

Controlling wave propagation — such as bending acoustic or elastic waves around objects to achieve invisibility — has emerged as a hallmark of metamaterial design. Invisibility cloaks, a long-held aspiration of humanity, have become feasible with the development of artificial metamaterials and transformation optics theory across various fields, including optics [1–4], acoustics [5,6], elastics [7], and thermodynamics [8]. For example, a three-dimensional elasto-mechanical cloak was demonstrated using pentamode metamaterials [9], which effectively conceals objects by mimicking the surrounding material's elastic response under deformation, validated by displacement field measurements.

On the other hand, an outstanding challenge of cloaking in solid mechanic is balancing both bulk and shear moduli in three dimensions. In elastomeric cellular solids, efforts have been made on switching of phononic properties by elastic instabilities under compression and reversible pattern transformations [10]. Critical hurdles remain as achieving precise control over both bulk and shear moduli in 3D solid mechanics, compounded by fabrication constraints and theoretical limitations in elastostatic invariance.

Due to their ease of processing, layered metamaterials are widely favored. By manipulating acoustic waves with layered metamaterials, acoustic invisibility cloaks have been demonstrated, which have wide implications like acoustic stealth and reverberation engineering. Besides invisibility cloaks, many applications using layered metamaterials have been proposed, such as non-destructive testing of Lithium-ion batteries [11], elastomeric bridge bearings [12], and geophysical exploration of the Earth's interior [13,14], etc. Compared with traditional solid and liquid materials in layered structure, soft metamaterials with their flexibility and formability show greater potential for wave control needed in complex environments. *A critical unresolved challenge is lack of theoretical framework connecting the microscale interface dynamics of soft matters(e.g., slip, adhesion) to macroscale wave phenomena.*

In this Letter, we reveal that soft layered structures exhibit three distinct wave characteristics at low P-wave frequencies corresponding to three types of interface effects: continuous, spring-like, and slip-like. We establish a unified theory for wave propagation in soft matter layered structures, resolving the interplay between interfacial conditions (quasistatic, resonant, and slip regimes) and effective dynamic properties. *By deriving analytical dispersion relations and homogenization limits, we demonstrate how soft interfaces enable tunable wave modes inaccessible to conventional rigid or fluidic composites.* This approach enhances our understanding of wave propagation in complex materials, with potential applications in various engineering fields.

Firstly, the wave characteristics of soft material layer are studied by establishing the dispersion relation of soft matters layered structure. The research object is alternate layered structure of two soft matters with thickness $h_1$ and $h_2$. Since the theory of propagation through layered media in the normal direction ($x_3$) is clear, we especially focus on the tangential direction ($x_1$). To perform this analysis, the potentials method is used to solve the Navier governing equation. Consequently, assuming in finite plane harmonic wave in the form

$$\phi = (A^+ \exp[i(k_p x_3)] + A^- \exp[-i(k_p x_3)]) \exp[i(k_1 x_1 - \omega t)],$$
$$\psi = (B^+ \exp[i(k_s x_3)] + B^- \exp[-i(k_s x_3)]) \exp[i(k_1 x_1 - \omega t)],$$
(1)

where $\omega$ is angular frequency, $k_1$ is the wave vector of $x_1$ direction, $k_p^2 = \omega^2 \rho / \kappa - k_1^2$ and $k_s^2 = \omega^2 \rho / \mu - k_1^2$. The displacements can be written in terms of the potentials as $\mathbf{u} = \nabla \phi + \nabla \times \psi$.

Applying the CIC, four homogeneous system of equations was funded; the determinant of the coefficient matrix of the equations must be zero to ensure that the solution is non-trivial. Details on the derivation process can also be found in the Supplemental Material. Therefore, we get the dispersion relationship as Eq.(S6).

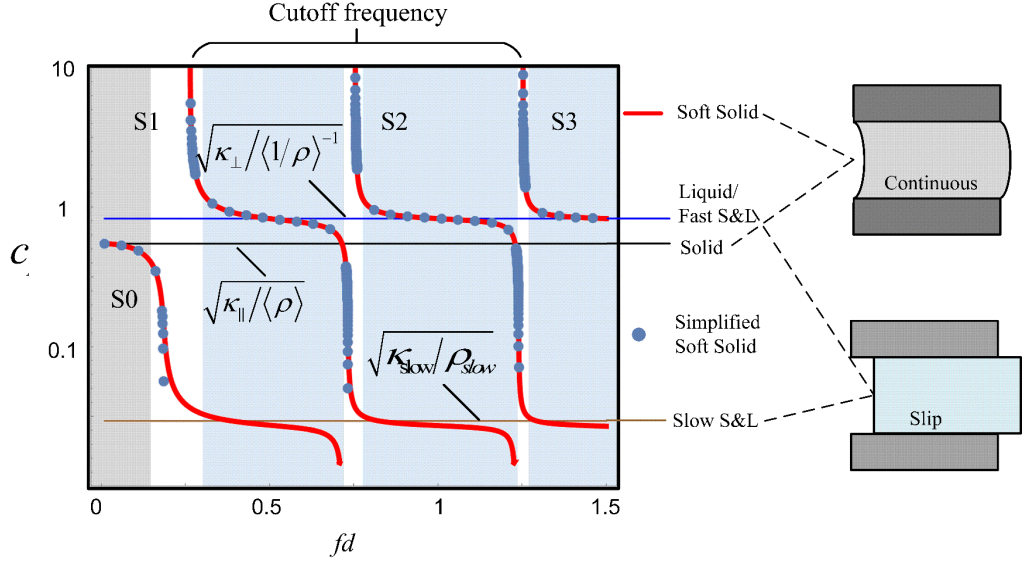

**Fig. 1** Dispersion relationship diagram for **solid (black solid line), liquid (blue line),** and **soft solid (red line)** laminated structure.

**Fig. 1** compares dispersive relationship of soft matter layered structures (shown in red) with three existing homogeneous models, solids (black), liquids (blue), and a combination of solid and liquid layered structures (blue and brown). Notably, the curves reveal differences in behavior and response to wave propagation, underscoring the unique properties of soft materials compared to traditional homogeneous models. The example parameters are set as $r_\rho = \rho^{(2)}/\rho^{(1)} = 7$, $r_\kappa = \kappa^{(2)}/\kappa^{(1)} = 8$, $\mu^{(2)}/\kappa^{(1)} = 10^{-2}$, $\mu^{(1)}/\kappa^{(1)} = 10^{-4}$, $r_h = h^{(2)}/h^{(1)} = 1$.

As shown in **Fig. 1**, in the low-frequency limit (when $fd$ is near zero), the dispersion relation of soft materials corresponds to the wave properties of solids, referred to as the Quasi-Static (QS) region (grey area).

As $fd$ increases, the dispersion relation shifts sharply in the Resonance Mode (RM) region (white area), reaching a critical point near 0.26, were unique for soft material。

Further increasing $fd$ to approximately 0.5 aligns the dispersion relation with those of the liquid and solid & liquid layered structure, located in the Slip Mode (SM) region (blue area). Additionally, it is observed that the RM and SM bands alternate, with the width of the white area narrowing as $fd$ continues to increase.

To highlight the role of the ratio $\mu/\kappa$ (where $\mu$ is the shear modulus and $\kappa$ is the bulk modulus), we present the distribution of the Quasi-Static (QS), Resonance Mode (RM), and Slip Mode (SM) regions as functions of the normalized frequency $fd^\kappa = \omega h^{(n)}/(2\pi \alpha^{(n)})$ in **Fig. 2** (a). Different colors in the figure represent distinct wave characteristics: the gray area indicates the QS region, the blue area represents the SM region, and the white area denotes the RM region. As illustrated in **Fig. 2** (a), the ratio of $\mu/\kappa$ significantly influences the wave characteristics of layered structures.

Initially, when the ratio of $\mu/\kappa$ is close to $10^{-1}$ (solid material), the QS region encompasses the entire P-wave low-frequency band. However, as this ratio decreases, the extent of the QS band shrinks, allowing the RM and SM bands to emerge sequentially. With a continued decrease in the ratio, both the RM and SM bands also narrow, leading to the emergence of new RM and SM bands. Ultimately, as the ratio approaches zero (indicating liquid material), the thickness of the QS region diminishes to zero, and the entire frequency range becomes dominated by the SM region. This transition underscores the complex interplay between material properties and wave behavior in layered structures.

The development of dispersion curves based on thickness contrast ratios is shown in **Fig. 2** (b) for examples with $r_h$ = 10, 1, and 0.1, represented by different colored solid lines. Overall, the variation of geometric parameters significantly influences structural parameters. As shown in **Fig. 2** (b), as $r_h$ increases, the cutoff frequency shifts to a lower frequency band; the velocities in the Quasi-Static (QS) region and the SM-fast region decrease. However, the velocity in the SM-slow region shows little variation with changes in $r_h$.

**Fig. 2** (c) *demonstrates the unique capability of soft matter layered structures to achieve simultaneous anisotropic density and modulus—a feat unattainable with conventional materials.* We plot the anisotropy ratio of modulus (vertical axis) against density anisotropy (horizontal axis) for a bilayer system where M1 is fixed (such as acrylic, with bulk modulus κ and density ρ) and M2 is varied. When layer M2 filled with a soft material (e.g., hydrogel), the system exhibits two distinct wave modes: S0 (solid line, quasistatic response) and S1 (dashed line, resonance-driven behavior). The black curves trace the full range of anisotropy combinations accessible by tuning the M1/M2 thickness ratio, forming a continuous design space (shaded region). In stark contrast, rigid solids (purple), liquids (blue), and gases (green) are confined to the axes, achieving only modulus (solids) or density (liquids/gases) anisotropy.

This dual tunability resolves a critical limitation in metamaterial design, where applications like acoustic cloaking or superlensing require independent control of effective density and stiffness. Soft matter layered systems thus establish a universal platform for engineering materials with on-demand anisotropic properties, critical for adaptive wave control in dynamic environments.

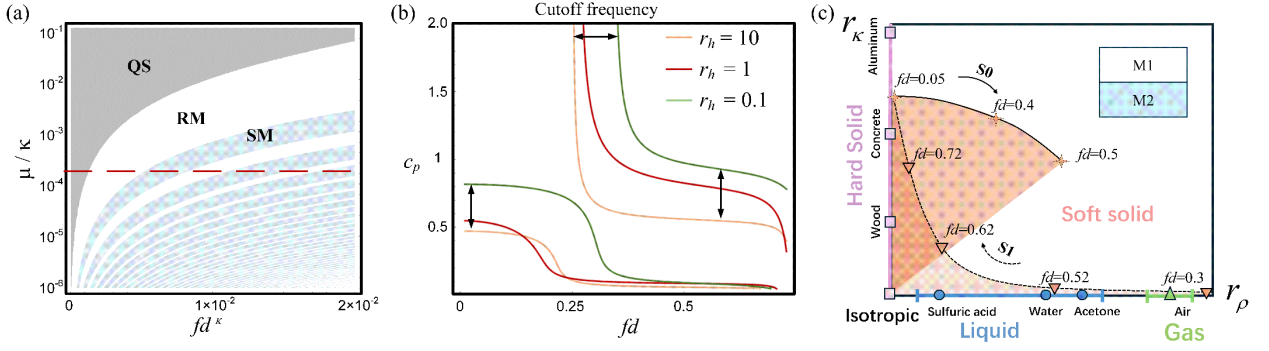

**Fig. 2** (a) The distribution of the Quasi-Static (QS), Resonance Mode (RM), and Slip Mode (SM) regions as functions of the normalized frequency $fd^{\kappa}$. (b) Dispersion diagrams showing normalized frequency response $fd = \omega h/(2\pi \beta_n)$ vs sound speed for three composites: $r_\rho = \rho^{(2)}/\rho^{(1)} = 7$, $r_\kappa = \kappa^{(2)}/\kappa^{(1)} = 8$, $\mu^{(2)}/\kappa^{(1)} = 10^{-2}$, $\mu^{(1)}/\kappa^{(1)} = 10^{-4}$, and $r_h = h^{(2)}/h^{(1)} = 10, 1, 0.1$. (c) Phase diagram of parameters contrast ratio of parallel vs vertical for soft, solid and liquid layered structure.

Next, the physical mechanism of the different wave characteristics of the soft matters layered structure is studied. As we all know, interface behavior plays a crucial role in the dynamic response of the layered structures and highlights the complex interactions that govern wave propagation in soft materials. Through comparing traditional solid and liquid materials with soft matters layered structure, our research identifies three distinct equivalent interface conditions, as depicted in **Fig. 3** (a). The first is the continuous interface condition. The second, known as the spring-like condition, arises at certain frequencies where the interface exhibits maximum standing amplitude. This results in significant shear stress and opposing average displacements across the interface, facilitating energy transfer between layers. Conversely, the slip-like condition occurs at other frequencies when the interface reaches a stagnation point. In this scenario, negligible shear stress is present, leading to differing average displacements across the interface. Generally, continuous interfaces exist in solid laminated structures, while slip interfaces are present in liquid laminated interfaces. However, soft materials exhibit three types of interfaces simultaneously: besides continuous and sliding interfaces consistent with solid-liquid laminated structures, there also exists a resonant interface unique to soft material laminated structures.

As the frequency parameter increases, both displacement and stress change, particularly at the interface of the two layers. Analyzing how wave structures vary with increasing frequency (*fd*) along a specific mode provides valuable insights. The normalized frequency $fd = \omega h^{(n)}/(2\pi \beta^{(n)})$ is reported where $h$ is the thickness and $\beta$ is the S-wave speed of soft layers. **Fig. 3** (b) illustrates the solutions for modes S0 and S1 at various *fd* values, with parameters such as $\rho^{(1)}/\rho^{(2)} = 7$, $h^{(1)}/h^{(2)} = 1$, $\kappa^{(1)}/\kappa^{(2)} = 8$, $\mu^{(1)}/\kappa^{(1)} = 10^{-2}$, and $\mu^{(2)}/\kappa^{(2)} = 10^{-4}$. **Fig. 3** (b) displays the structure (with deformation shown by solid lines), the displacement (represented as a density map), and the stress vector (indicated by arrows) at the interface of medium 1. A simplified model diagram adjacent to the main figure further elucidates the observed phenomena.

In addition to the quasi-static mode, four major modes of soft matter layered structures are presented in **Fig. 3** (b). At *fd* = 0.264, the interface exhibits maximum standing amplitude, indicating a spring-like condition. Here there are two distincting modes S0 and S1 arises from the type of spring. S0 corresponds to a stretch spring, while S1 represents a compressed spring. At *fd* = 0.5, the interface reaches a stagnation point, resulting in a slip-like condition. Here, two modes S0 and S1 exist., which relates to the direction of displacement. S0 moves in the opposite direction, while S1 moves in the same direction. These four modes and their simplified models reveal the reason why soft materials, although inherently solids, exhibit differences in wave propagation characteristics under continuous boundary conditions compared to solid laminated structures.

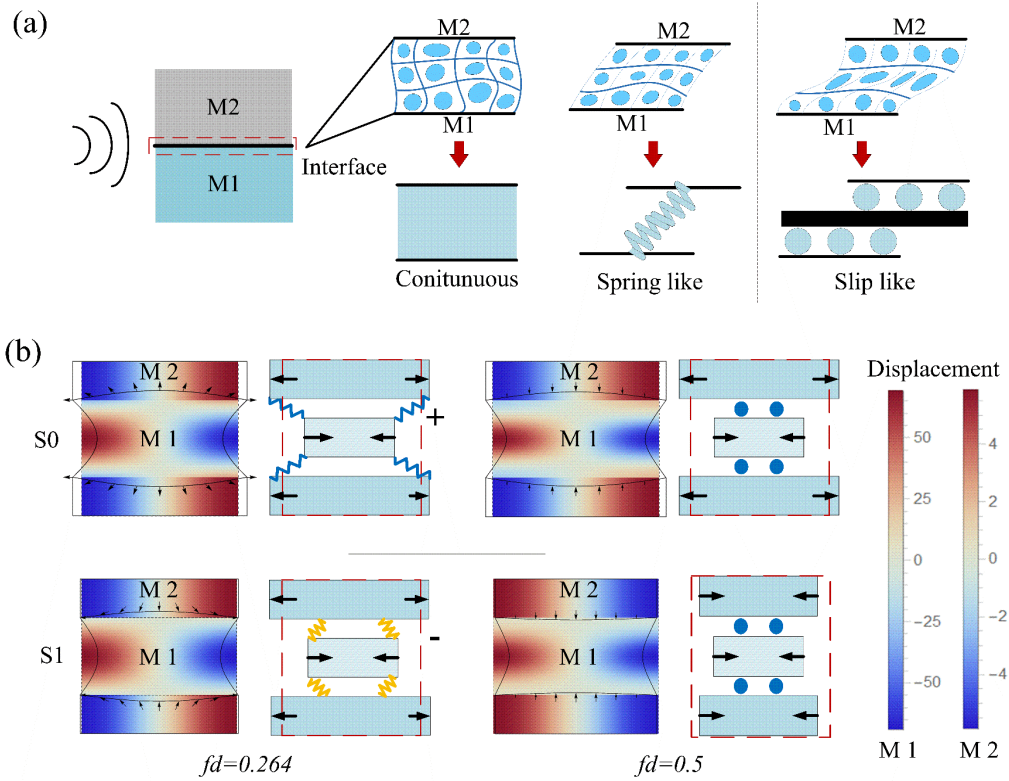

**Fig. 3** (a) Three distinct equivalent interface conditions in the soft matter layered structure. One is the continuous interface condition typical of solid-solid interfaces, while the other two—spring-like and slip-like conditions—are often found in resonant structures and liquid materials. (b) Four major modes of soft matter layered structures for various points on the S0 and S1 mode of an example with $\rho^{(1)}/\rho^{(2)} = 7$, $h^{(1)}/h^{(2)} = 1$, $\kappa^{(1)}/\kappa^{(2)} = 8$, $\mu^{(1)}/\kappa^{(1)} = 10^{-2}$, $\mu^{(2)}/\kappa^{(2)} = 10^{-4}$, showing the structure (without deformation, dashed line and with deformation, solid line), displacement (density map) and stress vector (arrow) on the mid layer interface. Their simplified model diagram is present on the right side.

Homogenization models are extensively applied in material science and engineering to predict the overall properties of composites. They simplify complex microstructures into a uniform continuum, allowing for material analysis and design with standard mechanics techniques. To this end, we study the homogenization model of soft matters layered structure. The wave characteristics of layered structure in the $x_3$ direction is clear in full frequency band, the equivalent density and modulus in the $x_3$ direction can be expressed $\rho_{eff3} = \langle \rho \rangle, \kappa_{eff3} = <1/\kappa>^{-1}$. In $x_1$ direction, base soft matter characteristic we have $\alpha^2 \gg \beta^2$. Therefore, we can simplify the result of dispersion curve Eq. (S6) and get the simplified result as follow

$$k_1^2 = \omega^2 \langle 1/\kappa \rangle \left[ \langle 1/\rho \rangle + \varphi_1 \varphi_2 \left(1 - \langle 1/\rho \rangle \langle \rho \rangle\right) / \langle \varphi \rho \rangle \right]^{-1}, \quad (2)$$

where $\varphi = \tan(2\pi fd)/(2\pi fd)$. Details on the derivation process can also be found in the Supplemental Material. From Eq.(2), the equivalent density and modulus in the $x_1$ direction can be derived as $\rho_{eff1} = \left[ \langle 1/\rho \rangle + \varphi_1 \varphi_2 \left(1 - \langle 1/\rho \rangle \langle \rho \rangle\right) / \langle \varphi \rho \rangle \right]^{-1}$, $\kappa_{eff1} = <1/\kappa>^{-1}$.

The simplified dispersion curve of Eq.(2) is shown in **Fig. 1** (a). As can be seen that the simplified result almost consistent with originate result. The error comes from the fact that the simplified results do not describe the solid & liquid (slow) mode.

The modulus of anisotropic modulus model can be degenerate into soft material and liquid. The density of soft material laminated structure can be degenerate into anisotropic density and modulus model. Therefore, we established the homogenization results that apply across the entire frequency spectrum and all three materials as follows:

$$E_{11} = 4<\mu> - 4<\mu^2/\kappa> + \left(1 - 2<\mu/\kappa>\right)^2 <1/\kappa>^{-1},$$
$$E_{33} = <1/\kappa>^{-1}, E_{13} = \left(1 - 2\mu/\kappa\right) <1/\kappa>^{-1}, E_{44} = <1/\mu>^{-1}, \quad (3)$$
$$\rho_{eff3}^{-1} = \langle \rho \rangle^{-1}, \rho_{eff1}^{-1} = \langle 1/\rho \rangle + \varphi_1 \varphi_2 \left(1 - \langle 1/\rho \rangle \langle \rho \rangle\right) / \langle \varphi \rho \rangle.$$

When the laminates consist of hard solid materials, the RM and SM bands disappear, $\rho_{eff1} = \langle \rho \rangle$. Therefore, the Eq. (3) can degenerate into the form of anisotropic modulus model [15]. When the laminates consist of liquid materials, the S-wave modulus vanish. Therefore, $E_{11} = E_{33} = <1/\kappa>^{-1}$, $\rho_{eff1} = \langle 1/\rho \rangle^{-1}$, and the Eq. (3) can degenerate into the form of anisotropic density model [16]. In conclusion, the proposed model Eq.(3) is more universal than the anisotropic density and anisotropic modulus models.

Here we proved the accuracy of the proposed unified model by analyzing the acoustic scattering field of a layered structure cloak. A cylindrical scatterer with a 25 cm radius is placed in a background medium. The scatterer is covered with a cloak. The cloak adopts four models, including one microstructure model (Mic) and three homogeneous models, respectively. The microstructure uses two layers of alternating linear elastic soft matter with a thickness of 5mm. The three homogeneous models correspond to unified model (UM) proposed in this Letter as Eq.(3), anisotropic density model (AD) and anisotropic modulus model (AM), respectively. The material parameters are detailed in the supplementary materials. A plane wave was excited to the left of the background field. Three frequencies of the plane wave were set *fd* = 0.05 in QS band, *fd* = 0.264 in RM band, *fd* = 0.5 in SM band. Detailed simulation conditions and material parameters can be found in the Supplemental Material.

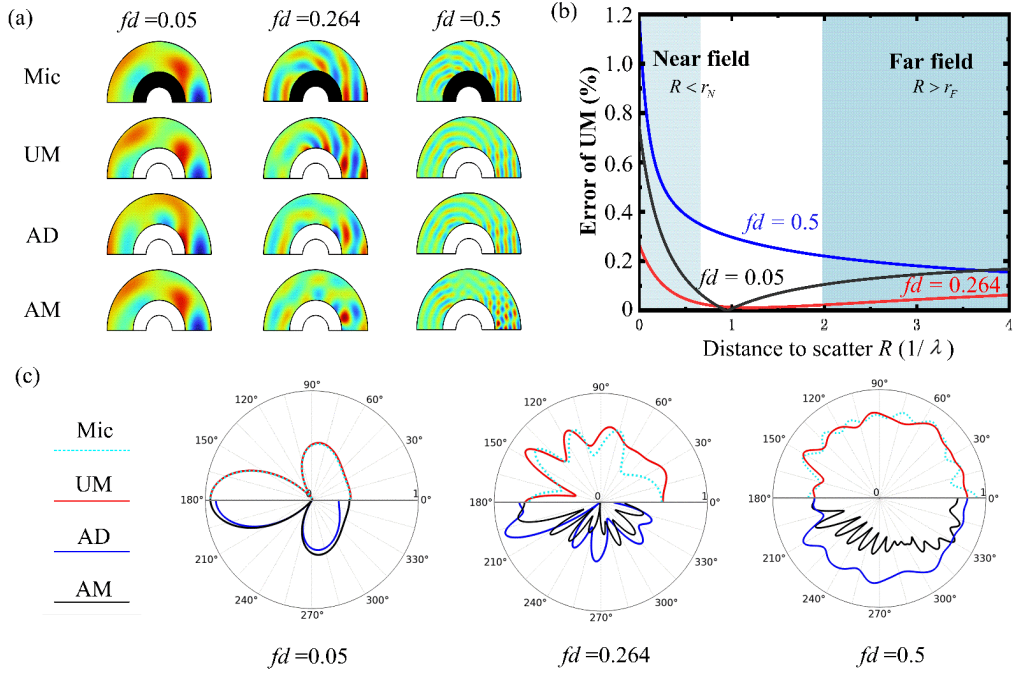

**Fig. 4** Scattered sound field diagram and boundary sound pressure distribution diagram. (a) The scattering field was calculated by using the microstructure model and homogeneous model. (b) Relationship between the error and distance to the cloak, showing the differences in the acoustic field between the microstructure model and the unified model in the QS, RM, and SM frequency bands. (c) Scattering sound pressure distribution at the outer boundary of the background.

The simulation results for the sound field are presented in **Fig. 4** (a), while the scattering sound pressure distribution at the outer boundary of the background is shown in **Fig. 4** (c) for four models across three frequency bands. It is observed that the AD model and the AM model exhibit similarities to the Mic model in specific frequency bands: the SM band and the QS band, respectively. The UM model introduced in this study demonstrates a superior ability to describe the wave characteristics of the layered structures across the entire frequency band. However, some discrepancies exist between the UM model and the Mic model in the RM and SM frequency bands. This discrepancy arises because two modes are present in the RM and SM frequency bands, and the UM model can only capture the main modes described in **Fig. 4**.

In addition, **Fig. 4** (b) illustrates the relationship between the error and distance to the cloak. It shows that the error in the near field (where $R < r_N = 0.62\sqrt{D^3/\lambda}$) is significantly larger than in the far field (where $R > r_F = 2D^2/\lambda$). This indicates that the models may perform less accurately in the near field, where complex wave interactions occur, while achieving better agreement in the far field. The consistency between the scattering field of the Mic model and that predicted by the UM throughout the full frequency band suggests that the unified model captures the essential features of the laminated structures more comprehensively.

In this study, we revealed three distinct wave characteristics of soft matter layered structures in the long wavelength limit of P-wave: quasistatic types, resonance types, and slip types. These wave characteristics are highly tunable based on material choices and layer thicknesses. We demonstrated that soft matters in layered composites can be simplified to liquids in the SM band, as the interface condition of the soft layer is similar to *SIC*. The wave structures in the RM and SM bands were analyzed, and a simplified model was established to explain the physical mechanisms behind the different wave characteristics. We also formulated a homogeneous model suitable for solid, liquid, and soft matter layered composites.

Our study demonstrates that soft-matter layered composites achieve simultaneous anisotropic density and modulus over a broad range—a critical advance for wave-manipulation technologies. Unlike conventional solids or fluids, the deformability and non-flow characteristics of soft matter enable adaptive, extreme anisotropic properties unattainable in rigid or fluidic systems. While rigid solids achieve modulus anisotropy and fluids density anisotropy, the dynamic interfacial behavior of soft matter bridges these regimes, offering unprecedented design freedom. Moreover, the compatibility of soft materials with scalable fabrication (e.g., layer-by-layer assembly) positions it as a practical solution for real-world applications, from wearable ultrasound devices to adaptive vibration dampers. These findings establish soft matter as a universal framework for next-generation metamaterials, paving the way to future exploration of active control and 3D architectures for multifunctional wave engineering.

The authors thank the value discussion with Yiwen Li and Nannan Jian throughout this study. This research was supported by the National Natural Science Foundation of China [Grants Nos. 11991031, 12272041, and 12202055] and the Aeronautical Science Foundation (ASFC-20230042072010). N. Fang thank the startup funding by the Jockey Club Trust (Grant number GSP181, N. Fang) STEM Lab of Scalable and Sustainable Photonic Manufacturing in the University of Hong Kong and from the Materials Innovation Institute for Life Sciences and Energy (MILES), HKU-SIRI in Shenzhen for supporting this manuscript.